\documentclass[aps,prl,twocolumn,groupedaddress,showpacs,floatfix,altaffilletter]{revtex4-1}
\usepackage{graphicx}
\usepackage{grffile}
\usepackage{amsmath}
\usepackage{amssymb}
\usepackage{bm}
\usepackage{color}
\usepackage[dvipsnames]{xcolor}
\usepackage{amsmath,amssymb,amsfonts}
\usepackage{epsfig}
\usepackage{times}
\usepackage[colorlinks,bookmarks=false,citecolor=blue,linkcolor=red,urlcolor=blue]{hyperref}
\usepackage{dsfont}

\newcommand{\fref}[1]{Fig.~\ref{#1}}
\newcommand{\eref}[1]{Eq.~(\ref{#1})}

\newcommand{\tref}[1]{Table~\ref{#1}}

\usepackage{fixfoot}

\begin{document}

\title{Superconducting Symmetries of Sr$_2$RuO$_4$ from First-Principles Electronic Structure}

\author{O.~Gingras$^1$\email{olivier.gingras.1@umontreal.ca}, R.~Nourafkan$^2$, A.-M.~S.~Tremblay$^{2,3}$, M.~C\^ot\'e$^1$}
\affiliation{$^1$D\'epartement de Physique and Regroupement Qu\'eb\'ecois sur les Mat\'eriaux de Pointe, Universit\'e de Montr\'eal, C. P. 6128, Succursale Centre-Ville, Montr\'eal, Qu\'ebec H3C 3J7, Canada}
\affiliation{$^2$D\'epartement de Physique, Institut quantique, Regroupement Qu\'eb\'ecois sur les Mat\'eriaux de Pointe, Universit\'e de Sherbrooke, Sherbrooke, Qu\'ebec, Canada}
\affiliation{$^3$Canadian Institute for Advanced Research, Toronto, Ontario, Canada M5G 1Z8}

\date{\today}

\begin{abstract}
Although correlated electronic-structure calculations explain very well the normal state of Sr$_2$RuO$_4$, its superconducting symmetry is still unknown. Here we construct the spin and charge fluctuation pairing interactions based on its correlated normal state.  Correlations significantly reduce ferromagnetic in favor of antiferromagnetic fluctuations and increase inter-orbital pairing. From the normal-state Eliashberg equations, we find spin-singlet $d$-wave pairing close to magnetic instabilities. Away from these instabilities, where charge fluctuations increase, we find two time-reversal symmetry-breaking spin-triplets: an odd-frequency $s$-wave, and a doubly-degenerate inter-orbital pairing between $d_{xy}$ and ($d_{yz},d_{xz}$).
\end{abstract}

\maketitle

Intensive experimental and theoretical studies have not yet yielded a definitive answer for the superconducting symmetry for Sr$_2$RuO$_4$ (SRO). Its similarities with $^3$He~\cite{rice_sr_1995} as well as experiments such as early temperature independent nuclear magnetic resonance across the critical temperature ($T_c$) ~\cite{ishida_spin-triplet_1998, ishida_spin_2015}, polarized neutron scattering~\cite{duffy_polarized-neutron_2000} and phase-sensitive tunneling experiments~\cite{liu_tunneling_2003, nelson_odd-parity_2004, liu_phase-sensitive-measurement_2010, liu_unconventional_2015} suggest a superconducting spin-triplet state.
Moreover, muon spin relaxation~\cite{luke_time-reversal_1998} along with polar Kerr effect~\cite{xia_high_2006} revealed the breaking of time-reversal symmetry in its superconducting state.
These two properties promptly lead to an assumption that the superconducting gap symmetry is chiral $p$-wave, implying a topological fully gapped state with $\bf d$-vector ${\bf d}=\hat{z}(k_x \pm ik_y)$~\cite{mackenzie_superconductivity_2003, maeno_evaluation_2012}.

On the other hand, low-temperature gapless excitations were found by various methods~\cite{nishizaki_changes_2000, tanatar_anisotropy_2001, izawa_superconducting_2001, lupien_ultrasound_2001}.
Residual thermal conductivity at very low temperature is difficult to reconcile with a nodeless state and rather supports a $d$-wave nodal state~\cite{hassinger_vertical_2017}.
Also, at the second critical magnetic field $H_{c2}$, the phase transition shows evidence of being first order and $H_{c2}$ is much lower than expected in a spin-triplet superconductor~\cite{yonezawa_specific-heat_2014, kittaka_sharp_2014}.
It suggests the existence of a pair-breaking mechanism similar to Pauli limiting, observed for spin-singlet superconductors.
Furthermore, while uniaxial strain experiments showed that the critical temperature could be enhanced by approaching van Hove singularities (vHS),  no signature of breaking of degeneracy between $k_x$ and $k_y$ was observed~\cite{steppke_strong_2017, watson_micron-scale_2018}. Moreover, very recently, the magnetic susceptibility was re-measured and found to drop below $T_c$ challenging a standard triplet pairing state~\cite{pustogow_constraints_2019, ishida_reduction_2019}.
These opposing observations make SRO one of the most mysterious modern theoretical puzzle in superconductivity and any step towards a better understanding could unravel important knowledge~\cite{mackenzie_even_2017}.

The multi-orbital nature of the superconductivity in SRO complicates the analysis.
Numerous studies have attempted to characterize the symmetry of SRO superconducting order parameter along with its dominant orbital host~\cite{baskaran_why_1996,mazin_ferromagnetic_1997, mazin_competitions_1999, kuwabara_spin-triplet_2000, kuroki_crib-shaped_2001, nomura_theory_2002, pavarini_first-principles_2006, chung_charge_2012, raghu_theory_2013, wang_theory_2013, scaffidi_pairing_2014, tsuchiizu_spin-triplet_2015, cobo_anisotropic_2016,huang_possible_2018, kaba_group-theoretic_2019}, yet this discussion remains open~\cite{mackenzie_even_2017}.
By contrast, the electronic structure of the normal state of SRO, including interaction-induced mass renormalizations, is well explained by first-principles approaches~\cite{mravlje_coherence-incoherence_2011}. It is thus desirable to use the machinery that describes well the normal state in order to address the unconventional superconductivity in SRO. Here we take a major step in that direction by finding out the leading superconducting instabilities from solving the Eliashberg equations starting from its correlated electronic structure.~\cite{Note1}

SRO is a single-layer perovskite, with the ruthenium atom in the center of a tetragonally elongated octahedron of oxygen atoms.
This configuration breaks the five-fold degeneracy of the $4d$ shell of ruthenium into $t_{2g}$ and $e_g$ states.
There are four electrons residing on the \textit{t$_{2g}$} subset, namely the $d_{xy}$, $d_{yz}$ and $d_{xz}$ orbitals, while the $e_g$ orbitals remain empty. Such a partially filled $4d$ subshell hosts relatively strong local  electronic interactions.

We thus start from a correlated electronic structure obtained using density-functional theory in the local density approximation plus dynamical mean-field theory (LDA+DMFT)~\cite{kotliar_electronic_2006}.
The LDA part of the electronic structure is computed using the projector augmented-wave pseudopotential~\cite{blochl_projector_1994, amadon_plane-wave_2008} implemented in ABINIT~\cite{torrent_implementation_2008,gonze_recent_2016}.
Although spin-orbit coupling in SRO affects some parts of the Fermi surface (FS)~\cite{haverkort_strong_2008, veenstra_spin-orbital_2014, zhang_fermi_2016, kim_spin-orbit_2018}, its effects on the spin and charge fluctuation spectra seems to be modest as we discuss in the supplemental material (SM)~\cite{NoteX}.
We neglect it at this stage.
We incorporate the correlation effects on $t_{2g}$ orbitals using the fully self-consistent LDA+DMFT~\cite{georges_dynamical_1996, kotliar_electronic_2006, lechermann_dynamical_2006} method with on-site Coulomb repulsion $U=2.3$~eV and Hund's coupling $J=0.4$~eV that are consistent with effective masses~\cite{mravlje_coherence-incoherence_2011, Note2}.

\begin{figure}
	\includegraphics[width=.98\linewidth]{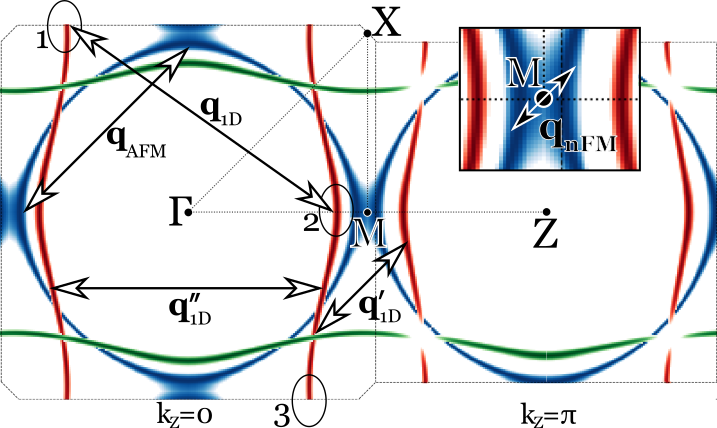}
	\caption{\label{fig:DMFT_FS} Partial in-plane spectral weight of Ru $t_{2g}$ orbitals on the FS  with $k_z=0$ (left) and $k_z = \pi$ (right) obtained from the LDA+DMFT calculation at $T=100$~K. Here $d_{xy}$ is blue, $d_{yz}$ is green and $d_{xz}$ is red. The two planes are next to each other because of the face-centered nature of the Brillouin zone. The principal nesting vectors are labeled. Important segments of the FS on the $d_{yz}$ orbital are encircled and numbered. Inset: nearly ferromagnetic vector ${\bm q}_{\text{nFM}}$ around the $M$ point.}\label{Spectral}
\end{figure}

Figure~\ref{fig:DMFT_FS} shows the LDA+DMFT in-plane partial spectral weights at the Fermi energy, $[\bm A({\bf k},\omega=0)]_{ll}$ with $l$ the orbitals $d_{xy}$, $d_{yz}$ and $d_{xz}$ illustrated by blue, green and red colors respectively. We interpret them as the FS of SRO, which consists of a cylindrical sheet ($\gamma$) and two quasi-one-dimensional (q1D) sheets ($\alpha$ and $\beta$): the $\gamma$ band is mainly derived from the $d_{xy}$ orbital and is close to a vHS, while the $\alpha$ and $\beta$ bands are mainly derived from the $d_{xz}$ and $d_{yz}$ orbitals. The main nesting vectors for different FS pockets are illustrated on \fref{fig:DMFT_FS}.

The effect of local correlations on the electronic structure, encoded in the self-energy $\bm \Sigma$, is to shift the non-interacting eigenenergies and to introduce a finite lifetime to quasiparticles. The orbital $l$'s quasiparticle renormalization factor $\bm Z_l$, where $\bm Z^{-1}_l \simeq 1 - \textrm{Im} [\bm\Sigma(i\omega_0)]_{ll}/\omega_0$ with $\omega_0$ the first Matsubara frequency,  is related to the effective mass enhancement via $m_l^*/m^{\text{LDA}}_l=\bm Z_l^{-1}$. We find $\bm Z_l^{-1} \sim 5.3, 3.8, 3.8$ for $d_{xy}, d_{yz}, d_{xz}$, consistent with quantum oscillation measurements~\cite{bergemann_quasi-two-dimensional_2003} and previous LDA+DMFT studies~\cite{mravlje_coherence-incoherence_2011}.

In spin and charge fluctuation mediated superconductivity, the Cooper pairing interaction is expressed in terms of spin and charge susceptibilities that measure the response to external magnetic or electric fields, respectively~\cite{esirgen_fluctuation_1998, senechal_theoretical_2004, nourafkan_nodal_2016, nourafkan_correlation-enhanced_2016}. They take large values at the wave vectors where the spin and charge fluctuations develop. The leading term is given by the bubble susceptibility
\begin{align}
	[\bm \chi^0_{ph}(Q)]_{l_1l_2;l_3l_4} & = - \frac{1}{N\beta} \sum_K \bm G_{K+Q,l_1l_3} \bm G_{K,l_4l_2},
\end{align}
where in LDA (LDA+DMFT) $\bm G_{K,l_1l_2}$ is the non-interacting (fully interacting) Green's function describing propagation of a particle from orbital $l_2$ to orbital $l_1$  with fermionic energy-momentum four vector $K\equiv (i\omega_m, {\bf k})$. The external bosonic energy-momentum four vector is $Q$ and its momentum $\bf q$ is called a nesting vector when the response is large because it nests different segments of the FS.

The propagator in an interacting system can be decomposed into coherent and incoherent parts: $\bm G \equiv \bm G^{coh}+\bm G^{incoh}$. Then the bubble susceptibility can be rewritten as the sum of two contributions: (i) one that comes from the product of the coherent (quasiparticle) parts of $\bm G$, (ii) another that comes from the scattering of the incoherent part with itself and with the coherent part~\cite{nourafkan_charge_2019}. The latter contribution is usually assumed smooth and featureless. The former contribution, that we will call quasiparticle, can be computed using ${\bm G}_{\text{QP}}={\bm Z}^{1/2}[i\omega_n {\bm 1}-{\bm H}_{\text{QP}}]^{-1}{\bm Z}^{1/2}$, where ${\bm H}_{\text{QP}}={\bm Z}^{1/2}[{\bm H}_{0}+{\rm Re}{\bm \Sigma}(0)-\mu{\bm 1}]{\bm Z}^{1/2}$~\cite{nourafkan_electric_2013}. It has been approximated in the literature by using the ARPES band structure in the expression  $[\bm \chi^{0,\text{QP}}_{ph}(Q)]_{l_1l_2;l_3l_4} \simeq \sqrt{\bm Z_{l_1}\bm Z_{l_2}\bm Z_{l_3}\bm Z_{l_4}} [\bm \chi^{0,\text{ARPES}}_{ph}(Q)]_{l_1l_2;l_3l_4}$~\cite{kreisel_orbital_2017}.

\begin{figure}
	\includegraphics[width=0.96\linewidth]{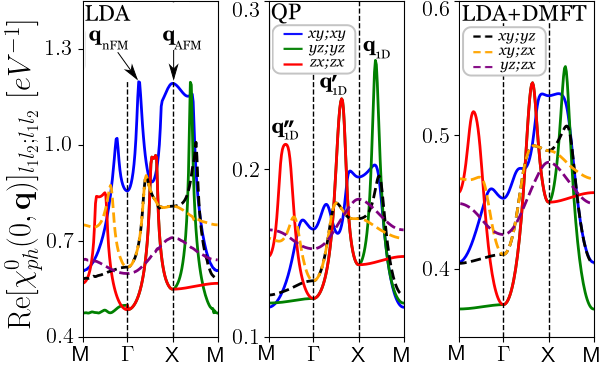}
	\caption{\label{fig:z-factors_ph} Comparison between LDA (left panel), QP (middle panel) and LDA+DMFT (right panel) components of the bubble p-h susceptibility $[\bm \chi^0_{ph}({\bf q}, \nu_n=0)]_{l_1l_2;l_1l_2}$ of SRO at $T = 100$~K. Each panel shows the intra-orbital (inter-orbital) components with full (dashed) lines. Dominant nesting vectors of \fref{fig:DMFT_FS} are labeled.}\label{Chiph}
\end{figure}

The important components of the bubble susceptibility in the particle-hole (p-h) channel, $\bm \chi^0_{ph}({\bf q},\nu_n=0)$, are plotted in \fref{fig:z-factors_ph} along a high-symmetry path. We highlight the effects of local electronic correlations by comparing the LDA, QP and LDA+DMFT bubble susceptibilities. Each panel shows the intra-orbital (inter-orbital) components with full (dashed) lines. The dominant peaks are labeled to correspond with the nesting wave vectors in \fref{fig:DMFT_FS}.

The cos-like shape of the q1D orbitals causes strong nesting at ${\bf q}_{1D} \sim (\pm\pi, \pm 2\pi/3), (\pm 2\pi/3, \pm\pi)$, as can be seen from purely intra-orbital $d_{xz}$ and $d_{yz}$ components. Other peaks benefiting from this q1D nature are ${\bf q}^{\prime\prime}_{1D}$ near the $M$ point and ${\bf q}^{\prime}_{1D} \sim (\pm 2\pi/3, \pm 2\pi/3)$ that corresponds to the neutron scattering observations in Ref.~\citenum{sidis_evidence_1999} and previous LDA+DMFT calculations~\cite{boehnke_multi-orbital_2018}.

The $d_{xy}$ intra-orbital component exhibits a wide plateau around the antiferromagnetic (AFM) nesting vector ${\bf q}_{\rm AFM} = (\pm\pi, \pm\pi)$ that connects states near vHSs.  Moreover, these states also induce strong nearly ferromagnetic (nFM) fluctuations at small ${\bf q}_{\rm nFM}$. The corresponding AFM and nFM instabilities compete in LDA, but in QP and LDA+DMFT the nFM peak is strongly suppressed by electron correlations, in agreement with inelastic neutron-scattering observations~\cite{iida_inelastic_2011}. Therefore, correlation effects reduce the tendency towards ferromagnetic ordering.

In the LDA calculation, the dominant component of $\bm \chi^0_{ph}$ is the $d_{xy}$ intra-orbital component. The $d_{xy}$ orbital has the strongest mass enhancement, or smallest $Z$, so in QP the $d_{xy}$ intra-orbital component is strongly reduced. Hence in QP, the dominant components are those from $d_{xz}$ and $d_{yz}$. In the full LDA+DMFT calculation, however, all orbitals have comparable susceptibilities as can be seen from the right panel of \fref{fig:z-factors_ph}. 

\begin{figure}
	\includegraphics[width=.92\linewidth]{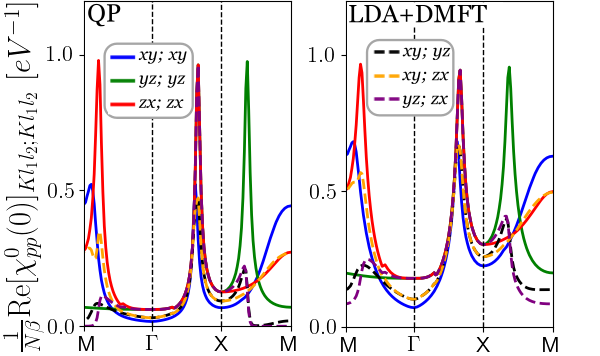}
	\caption{\label{fig:z-factors_pp}Real part of the bubble pairing susceptibility at the lowest fermionic frequency along a high-symmetry path for QP (left panel) and LDA+DMFT (right panel). }\label{Chipp}
\end{figure}

The Cooper pairing susceptibility, to lowest order, is obtained from $\left[{\bm \chi}^{0}_{pp}(0)\right]_{K,l_1 l_2;K',l_3 l_4}=(N\beta/2)\bm G_{K,l_1l_3}\bm G_{-K,l_2l_4}\delta_{K,K'}$. \fref{Chipp} shows the real part of several components of $(1/N\beta){\bm \chi}^{0}_{pp}(0)$ at the lowest fermionic frequency. The intra-orbital components (full lines) are purely real and show relatively sharp peaks at the position of FSs. In the LDA (not shown) and QP calculations, the peak heights are proportional to the corresponding orbital weight on the FSs and inversely proportional to the square of the Fermi velocity~\cite{nourafkan_nodal_2016}. They get narrower when reducing temperature, implying that only electrons on FSs contribute to pairing. The correlation effects broaden these peaks, so electrons away from the Fermi level can contribute to pairing. Moreover, the inter-orbital components (dashed lines) are considerably enhanced by correlations, hence electrons on different orbitals can form Cooper pairs.

In an interacting system, a propagating particle excites p-h pairs. This modifies the propagator and hence the bubble susceptibility. These corrections are captured in the LDA+DMFT calculation. However p-h excitations can be absorbed by a propagating hole, leading to a correction that is not included in the bubble susceptibility.  This correction can be accounted for using the Bethe-Salpeter equation as
\begin{equation}
	\bm \chi^{d/m}_{ph}(Q) = \frac{\bm \chi^0_{ph}(Q)}{{\bm 1} +/- \bm \Gamma^{d/m}_{ph}\bm \chi^0_{ph}(Q)},\label{Bethe-Salpeter}
\end{equation}
where $\bm \Gamma^{d/m}$ denotes p-h irreducible amplitudes for all scattering processes in density and magnetic channels. Here, we use the random phase approximation (RPA), which approximates the irreducible vertex functions with uniform and static (momentum and frequency independent) ones, as was done to study superconductivity in the cuprates~\cite{esirgen_fluctuation_1998, senechal_theoretical_2004} and the iron-based superconductors~\cite{graser_near-degeneracy_2009, kuroki_pnictogen_2009, yanagi_two_2010, nourafkan_nodal_2016, nourafkan_correlation-enhanced_2016, bekaert_advanced_2018}. In this approximation, the irreducible vertex functions are parametrized with screened interaction parameters $U_s$ and $J_s$ which are different from the bare interaction parameters entering in the DMFT calculations~\cite{NoteX, nourafkan_correlation-enhanced_2016}.
As one can see from \eref{Bethe-Salpeter} an instability in the magnetic (charge) channel occurs once the largest eigenvalue of $\bm \Gamma^{m}_{ph}\bm \chi^0_{ph}(Q)$ ($-\bm \Gamma^{d}_{ph}\bm \chi^0_{ph}(Q)$) reaches unity.  Hence, the Stoner factors, denoted $S^{m(d)}$ and defined by the largest eigenvalue, measure proximity to an instability.

The Cooper pairing interaction can be written in term of these susceptibilities, \eref{Bethe-Salpeter}, as explained in SM~\cite{NoteX}. In the singlet (triplet) channel, the charge and spin fluctuations compete (cooperate) in the effective pairing interaction. Their magnitude is determined by the $J_s/U_s$ ratio. At $J_s=0$, the intra- and inter-orbital interaction strengths are equal, hence, there is no energy difference between electronic configurations with electrons in the same orbital or in different ones. This leads to large charge fluctuations. On the other hand, a finite $J_s$ decreases the charge fluctuations.

Before showing results of detailed calculations in \fref{fig:leading_eigs}, it is instructive to analyze the pairing interaction to understand what pairing symmetries are most natural. Since RPA vertex functions are momentum independent, the RPA dressed susceptibilities share peak structures similar to the bubble ones, but with changes in relative magnitude that can become important close to instabilities. Nevertheless, by inspecting the bubble susceptibilities one can gain insights about the possible pairing symmetries. Here we consider $d_{xy}$ and $(d_{xz},d_{yz})$ separately and discuss their possible gap symmetries in the spin-singlet and triplet channels. We begin our preliminary discussion with the intra-orbital pairing and then comment on the possibility of inter-orbital pairing.

In the singlet channel, the overall pairing interaction is repulsive, requiring a sign changing gap symmetry.
The dominant nesting vectors of the $d_{xy}$ orbital corresponds to the plateau near $\bf{q}_{\rm AFM}$. Although the nesting condition seems poor, it is the leading wave vector for fluctuations because it connects the FS states around vHSs.  Hence, a gap function with large superconducting gap value at the FS patches around vHSs would have a much lower energy than the normal state, making a gap function with $d$-wave $\cos k_x - \cos k_y$ symmetry a prime candidate. In the LDA calculation, the nFM peak promotes degenerate gap functions with $p$-wave $\sin k_x$ or $\sin k_y$ symmetries. To satisfy Pauli's principle, these spin-singlet intra-orbital odd-parity states are odd in frequency and therefore have a vanishing equal-time order parameter ~\cite{NoteX, schrieffer_odd_1994, linder_odd-frequency_2017}. It was shown that odd-frequency states are thermodynamically stable and exhibit ordinary Meissner effect~\cite{linder_odd-frequency_2017, solenov_thermodynamical_2009, kusunose_possible_2011}.
Comparing to $\cos k_x - \cos k_y$ symmetry, the latter should be sub-leading gap symmetries due to nearby vHS. They become even less likely in LDA+DMFT because the nFM peak is suppressed by interaction.

The $d_{xz}$ ($d_{yz}$) q1D orbitals have dominant nesting vectors at ${\bf q}_{1D}$, which is compatible with a singlet gap of the form $\cos k_x$ ($\cos k_y$) with nodes on the FS near $k_x = \pm \pi/2$ ($k_y = \pm \pi/2$).
If the two orbitals are out of phase, then the resulting gap function has $d$-wave $\cos k_x - \cos k_y$ symmetry, while an in-phase gap function would rather have an$s^{\pm}$ symmetry~\cite{NoteX}.

Coupling all the orbitals together, the most probable singlet gap symmetry has $\cos k_x - \cos k_y$ $d$-wave symmetry.

In the triplet channel, the pairing interaction has both attractive and repulsive components, involving particle and hole momenta ( $K\uparrow, K^{\prime}\downarrow$) (for the $S^z=0$ case). The attractive (repulsive) parts are maximum when the transferred momentum ${\bf k}^{\prime}-{\bf k}$ (${\bf k}^{\prime}+{\bf k}$) is equal to a nesting vector and $\omega_m^{\prime}=\omega_m$  ($-\omega_m$)~\cite{NoteX}. For the $d_{xy}$ orbital, both attractive and repulsive components pair the same states because for ${\bf k}=(\pi,0)$ and ${\bf k}^{\prime}=(0,\pi)$ both ${\bf k}^{\prime}\pm{\bf k}$ correspond to ${\bf q}_{\rm AFM}$, which is where $d_{xy}$'s intra-orbital susceptibility peaks (see \fref{Chiph}).
For an even-frequency gap, these components therefore compete with each other leading to an overall suppression of Cooper pairing. This also can be seen differently. In the even-frequency triplet channel, the intra-orbital gap function has odd-parity, i.e., $\bm \Delta(-{\bf k})=-\bm \Delta({\bf k})$. It is maximum at the momentum position of the vHSs. However, as can be seen form \fref{Spectral}, the vHSs momenta are almost time-reversal invariant momenta (TRIM). A TRIM satisfies ${\bf k}_{\text{TRIM}} = -{\bf k}_{\text{TRIM}} + {\bf b}$ with $\bf b$ a reciprocal lattice vector, which implies $\bm\Delta(-{\bf k}_{\text{TRIM}})=\bm\Delta({\bf k}_{\text{TRIM}})$~\cite{yao_topological_2015}.  This contradicts the odd-parity relation. Hence, electrons on the $d_{xy}$ orbital would not condense in an odd-parity pairing channel.
Another possibility is an intra-orbital odd-frequency state, for which attractive and repulsive components of the interaction cooperate leading to an enhancement of Cooper pairing. Benefiting from vHSs, a gap function with $s$-wave symmetry $\Delta_0+\Delta_1(\cos k_x + \cos k_y)$ is preferred as suggested for q1D systems~\cite{shigeta_competition_2011}.

For $(d_{xz}, d_{yz})$ orbitals, attractive and repulsive parts of the interaction pair different states. For example, for $d_{xz}$ (red curves in \fref{Spectral} with numbers labelling encircled states), the attractive part pairs states $1$ and $2$ on two FS branches. On the figures, these states are connected with ${\bf q}_{1D}$. On the other hand, the dominant repulsive part pairs states $2$ and $3$ on the same FS branch. The resulting even-frequency gap function has two nodes on each FS branch and the two FS branches are out of phase. Furthermore, if gap functions for $d_{xz}$ and $d_{yz}$ orbitals are out of phase by $\pi/2$ then the resulting gap symmetry is what is predicted by Ref.~\citenum{raghu_hidden_2010}.  If the gap is odd in frequency, the requirement of odd-parity is lifted and the $\Delta_0 + \Delta_1 \cos k_x$ ($\Delta_0 + \Delta_1 \cos k_y$) symmetry for the $d_{yz}$ ($d_{xz}$) orbital is preferred. The gap function is maximum at $k_y = (0, \pm \pi)$ ($k_x = (0, \pm \pi))$ where there are more states.

With all orbitals included, the most probable gap symmetry in the triplet channel has an odd-frequency $\Delta_0+\Delta_1(\cos k_x + \cos k_y)$ extended $s$-wave symmetry.

All intra-orbital pairing described so far are non-local. Considering the inter-orbital pairing, a \emph{local} pairing mechanism becomes possible. Even in the presence of electronic repulsion, such pairing arises from Hund's coupling and promotes spin-triplet inter-orbital states that are odd under exchange of orbitals with an almost uniform momentum dependence (see SM for a detailed analysis of the pairing interaction components promoting inter-orbital Cooper pairing). Such states have been discussed in the context of SRO~\cite{puetter_identifying_2012}.

\begin{figure}
	\includegraphics[width=.75\linewidth]{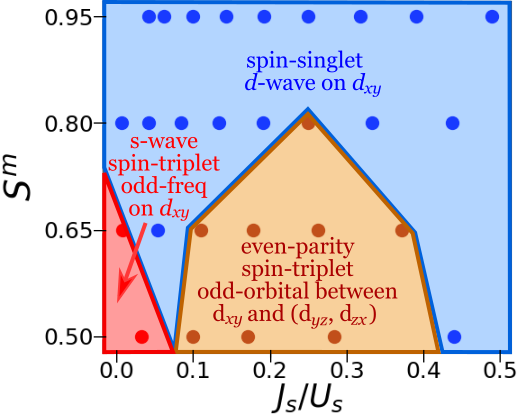}
	\caption{\label{fig:leading_eigs}Phase diagram of the leading superconducting instabilities. A lower $J_s/U_s$ implies more charge fluctuations, while the magnetic Stoner factor $S^m$ quantifies the proximity to a magnetic instability.}
\end{figure}

To tell apart all these possibilities, we performed an unbiased calculation by searching the leading eigenvalues and corresponding gap functions of the linearized normal-state Eliashberg equation for different combinations of $J_s$ and $U_s$~\cite{NoteX}.
Since the precise values of these parameters are unknown, we work with physically relevant ranges for $S^m$ and $J_s/U_s$.
We focus on $S^m$ between $0.5$ and $0.95$ as SRO is considered to be in the vicinity of a magnetic instability because $3$\% of manganese doping is enough to reach a magnetically ordered phase~\cite{ortmann_competition_2013}.
For $J_s/U_s$, various values appear in the literature, ranging between $0$ and $0.5$~\cite{raghu_theory_2013, wang_theory_2013, scaffidi_pairing_2014, tsuchiizu_spin-triplet_2015}.


Although we calculated several leading eigenvectors~\cite{NoteX}, \fref{fig:leading_eigs} shows only the leading gap symmetry for various points in parameters space.
A feature present in all of them is the importance of the $d_{xy}$ orbital as a host, consistent with experiments~\cite{deguchi_determination_2004, kunkemoller_absence_2017}.
In the vicinity of a magnetic transition, the system is dominated by a spin-singlet $d_{x^2-y^2}$-wave state hosted by the $d_{xy}$ orbital.
At smaller $S^m$, charge fluctuations are important and compete against (collaborate with) spin fluctuations in the singlet (triplet) channel.
Thus the $d$-wave state is suppressed and two spin-triplet states become dominant: (i) an odd in frequency intra-orbital $s$-wave state hosted by the $d_{xy}$ orbital and (ii) a set of degenerate even-parity, odd-orbital gap functions that pair states between the $d_{xy}$ and ($d_{xz}, d_{yz}$) orbitals.

Both of these symmetries are promising candidates to explain experimental results in SRO. The odd-frequency gap function can lead to an intrinsic Kerr effect as observed in SRO~\cite{komendova_odd-frequency_2017}. Also, the vanishing of its order parameter at zero frequency could mimic the presence of nodes since the building up of its gap away from the FS changes the quasiparticle spectrum~\cite{abrahams_properties_1995},
an effect that could be interpreted as the V-shape density of state observed in tunneling spectroscopy studies~\cite{firmo_evidence_2013}. Measurements of the zero-bias tunneling under magnetic fields, proposed as a fingerprint for odd-frequency in MgB$_2$~\cite{aperis_ab_2015}, would be relevant in the present context.
In the odd-orbital pairing, the two degenerate order parameters $\Delta_1,\Delta_2$ can either form nematic states that break $C_4$ or form chiral states that break time-reversal symmetry as $\Delta_1 \pm i\Delta_2$.

In summary, we performed a systematic search for superconducting states of SRO from a LDA+DMFT electronic structure combined with static vertex functions (RPA). Using spin and charge fluctuation mediated pairing, we found that in proximity to a magnetic instability, the spin-singlet $d$-wave state is favored by AFM fluctuations.
Further away from the magnetic instability, charge fluctuations become sizeable, promoting two spin-triplet states: an odd-frequency $s$-wave state and a doubly degenerate odd-orbital state that pairs electrons between $d_{xy}$ and ($d_{yz}, d_{xz}$).
Both states are interesting candidates for superconductivity in SRO. Consequences on physical observables of these two pairing states should be studied in details.

\begin{acknowledgments}
This work has been supported by the Canada First Research Excellence Fund, the Fonds de Recherche du Qu\'ebec---Nature et Technologie (FRQNT), the Natural Sciences and Engineering Research Council of Canada (NSERC) under grants RGPIN-2014-04584 and RGPIN-2016-06666, and by the Research Chair in the Theory of Quantum Materials (AMST). Simulations were performed on computers provided by the Canadian Foundation for Innovation, the Minist\`ere de l'\'Education des Loisirs et du Sport (Qu\'ebec), Calcul Qu\'ebec, and Compute Canada.
\end{acknowledgments}

%

\pagebreak

\begin{center}
\textbf{\large Supplemental Materials: Superconducting Symmetries of Sr$_2$RuO$_4$ from First-Principles Electronic Structure}
\end{center}

\setcounter{equation}{0}
\setcounter{figure}{0}
\setcounter{table}{0}
\setcounter{page}{1}
\makeatletter
\renewcommand{\theequation}{S\arabic{equation}}
\renewcommand{\thefigure}{S\arabic{figure}}
\renewcommand{\bibnumfmt}[1]{[S#1]}
\renewcommand{\citenumfont}[1]{S#1}

In this Supplemental Material, we show in the first section the non-interacting LDA bands, with and without spin-orbit coupling (SOC). This allows us to argue that SOC does not substantially modify the susceptibilities in either the particle-particle or particle-hole channels. The second section recalls the Eliashberg equation and pairing vertex entering the calculations. The third section explains the difference between the present and the standard uses of the Eliashberg theory. The fourth section discusses the properties of the various gap functions. The fifth section presents the leading eigenvalues and the symmetry of the corresponding eigenvectors for various Stoner factors. Finally, the last section presents a detailed analysis of the pairing interaction components promoting inter-orbital Cooper pairing.

\section{Spin-orbit Coupling}\label{Sec:Bands}
\begin{figure}[b]
	\includegraphics[width=0.95\linewidth]{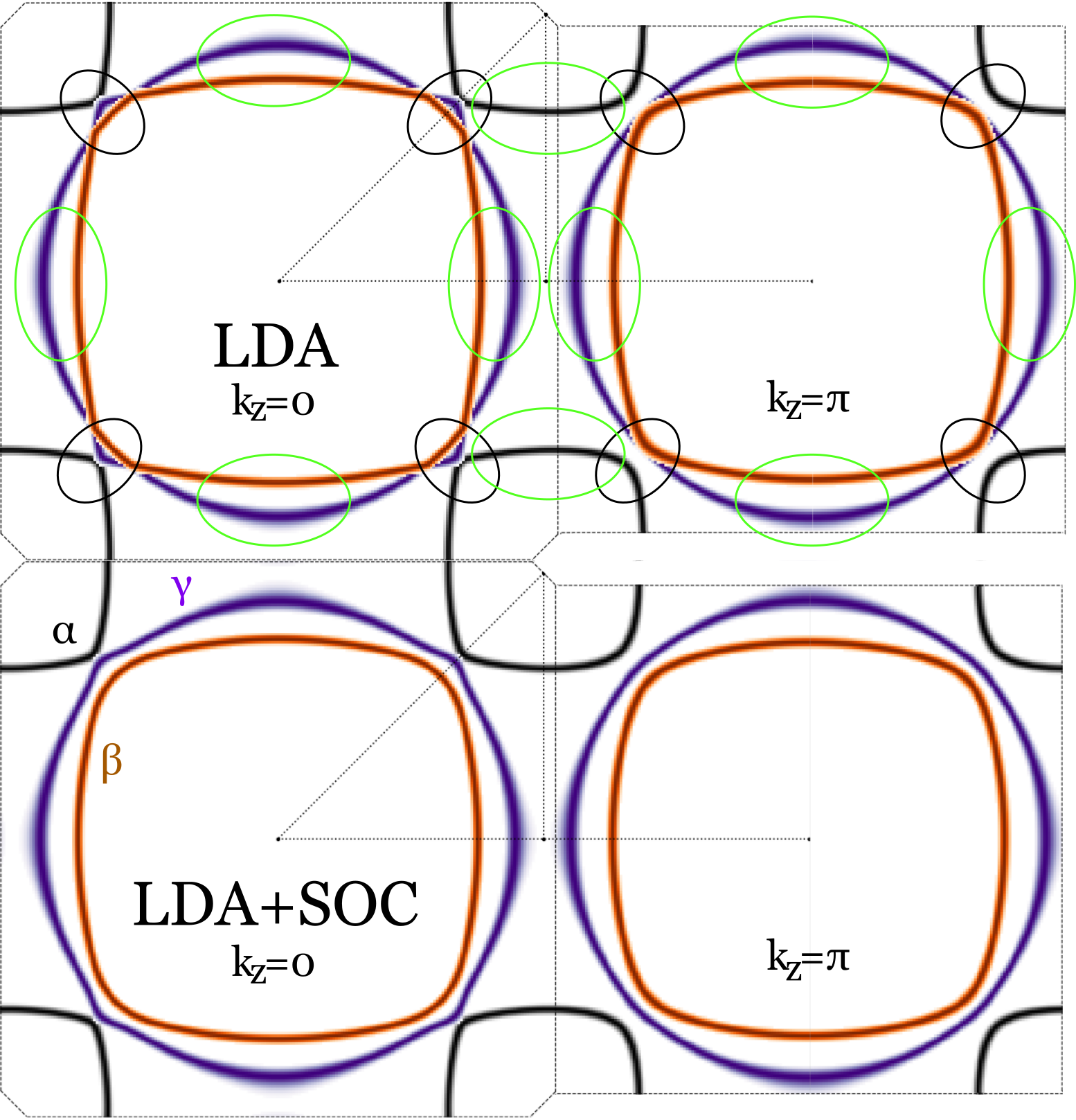}
	\caption{\label{fig:LDA_SOC_FS}Fermi surfaces in a) LDA and b) LDA+SOC at $k_z=0$ (left) and $k_z=\pi/c$ (right). Highly degenerate regions are encircled in black and important nesting regions in light green.}
\end{figure}

Spin-orbit coupling (SOC) was neglected in this work because it necessitates a more general formulation of the pairing vertex, yet to be developed.
Moreover, we believe that including it would only generate modest consequences on the results of this work and this section explains why.

\begin{figure}
	\includegraphics[width=\linewidth]{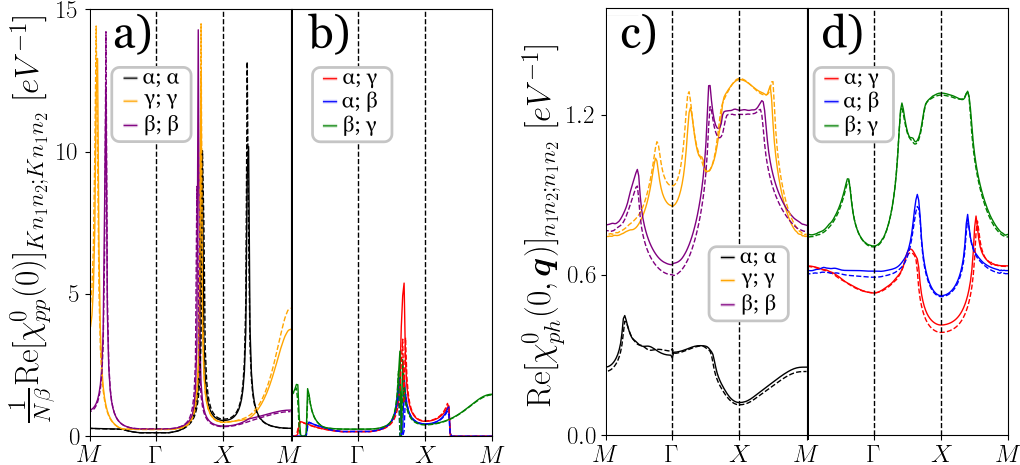}
	\caption{\label{fig:chi_SOC_vs_LDA} Bubble susceptibilities in the band basis for both LDA (full lines) and LDA+SOC (dashed lines) a)-b) Dominant components of the particle-particle susceptibility at the lowest fermionic frequency and c)-d) dominant particle-hole susceptibility. Intra-band components are in a)-c) and inter-band components are in b)-d), respectively.}
\end{figure}

\fref{fig:LDA_SOC_FS} shows all components of the partial spectral weight $[\bm A(\textbf{k},\omega=0)]_{nn}$ in the band basis that cross the Fermi surface. There are three such bands, namely $\alpha$, $\beta$ and $\gamma$. $\bm A_{nn}$ are interpreted as the Fermi surface (FS).
The FS in the top panels was obtained from density-functional theory in the local density approximation (LDA) and the lower panel ones from LDA with SOC (LDA+SOC).
The main effect of including SOC is to mix the orbital content near degenerate points, circled in black on \fref{fig:LDA_SOC_FS}.
The degeneracy is lifted and thus the FS slightly changes, with impacts on the band nesting, the orbital character and the spin character of the resulting bands~\cite{S_haverkort_strong_2008, S_veenstra_spin-orbital_2014, S_zhang_fermi_2016}.

However, the region contributing most to spin-fluctuations, circled in light green on \fref{fig:LDA_SOC_FS}, are away from these band-mixing points. 
It suggests that the bare particle-hole (p-h) susceptibility should not change significantly between LDA and LDA+SOC.
Figure \ref{fig:chi_SOC_vs_LDA} compares, between LDA (full lines) and LDA+SOC (dashed lines), the intra-band (a and c) and inter-band (b and d) components of the bubble susceptibilities in the particle-particle (p-p) (a and b) and particle-hole (p-h) (c and d) channels.

In the p-p channel, the main changes occurs between $\Gamma$ and $X$, which are precisely the nearly degenerate points encircled in black on \fref{fig:LDA_SOC_FS}.
The intra-orbital components illustrates the separation of the bands. Consequently, the inter-orbital peak between the $\gamma$ and ($\alpha$, $\beta$) bands is split.

In the p-h channel, there is a slight difference all around the Brillouin zone between LDA and LDA+SOC because of various changes in the FS.
Nevertheless, all the leading peaks keep similar positions, even though some are a bit suppressed.
Thus no significant changes in the gap functions are expected. 

The effect of SOC was argued not to affect the correlation-induced renormalizations incorporated through dynamical mean-field theory (DMFT)~\cite{S_kim_spin-orbit_2018}.
The previous conclusions should thus remain true in the LDA+SOC+DMFT framework, although further studies should be made.

\section{Bethe-Salpeter equation and Random phase approximation}\label{Sec:BetheSalpeter}

\begin{figure}
	\includegraphics[width=\linewidth]{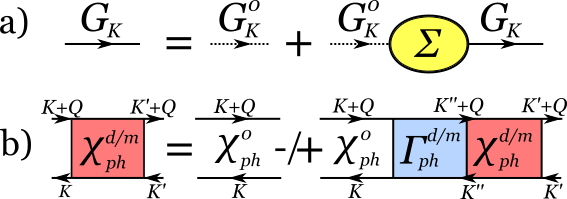}
	\caption{\label{fig:dressed_G_chi}a) The fully interacting Green's function $\bm G$ in constructed from the bare Green's function $\bm G^0$ and the self-energy $\bm \Sigma$. b) The dressed particle-hole susceptibility $\bm \chi_{ph}$ can be decomposed into the bubble susceptibility $\bm \chi^0_{ph}$ and the vertex correction part. The bubble susceptibility describes independent, but interaction-renormalized, propagation of a particle-hole (p-h) excitation and is obtained from fully interacting Green's function $\bm G$.}
\end{figure}

In an interacting system, the propagating particle-hole excitations interact with their environment through their self-energy cloud and with each other by exchanging multiple real or virtual p-h excitations. According to the Bethe-Salpeter equation, the dressed susceptibilities can be decomposed into the bubble susceptibility and the vertex corrections as shown in the \fref{fig:dressed_G_chi}~b). 
The propagator lines in this equation are described by the fully interacting Green's function obtained from LDA+DMFT which includes the self-energy, as shown in \fref{fig:dressed_G_chi}~a). We use the bare parameters $U=2.3$ and $J=0.4$ in our LDA+DMFT calculation. These parameters are the ones used in previous LDA+DMFT calculations that concerned effective masses and spectral weight. The bubble susceptibilities $\bm \chi^0_{ph/pp}$ introduced in the main text correspond to the lowest perturbation order of the response functions. We construct them from the fully interacting Green's function.

In the p-h channel, the dressed susceptibility includes the vertex correction part as well. Calculating this part requires the irreducible vertex function $\bm \Gamma_{ph}$, as shown in \fref{fig:dressed_G_chi}~b). Here, we approximate this vertex using the local and static Coulomb vertex $\bm \Gamma^{0,d/m}_{ph}$, obtained from the rotationally invariant Slater-Kanamori Hamiltonian. It yields
\begin{equation}
	\label{eq:irr_ph_vertex}
	[\bm\Gamma^{0,d(m)}_{ph}]_{l_1l_2;l_3l_4} = \left\{ \begin{array}{l l l}
		U_s			& (U_s)		& l_1 = l_2 = l_3 = l_4 \\
		-U_s'+2J_s	& (U_s')		& l_1 = l_3 \neq l_2  = l_4 \\
		2U_s'-J_s		& (J_s)		& l_1 = l_2 \neq l_3 = l_4 \\
		J_s			& (J_s)		& l_1 = l_4 \neq l_2 = l_3 \\
		0			& (0)			& \text{otherwise}
	\end{array} \right.
\end{equation}
where $U_s$ ($U'_s$) is the local and static screened intra-orbital (inter-orbital) on-site Coulomb repulsion. Because of rotational invariance, one has $U'_s = U_s - 2J_s$ and $J_s$ is the Hund's coupling. This approximation corresponds to the lowest order of parquet equations for the vertex and to the random phase approximation for the density and magnetic susceptibilities. The dressing of spin and charge fluctuations is performed using Eq. (2) of the main text.

The parameters $U_s$ and $J_s$ are different from the bare parameters ($U$ and $J$) due to screening effects. The values of $U_s$ and $J_s$ are unknown. 
By studying the effect of the Stoner factors, we can tune the distance from a magnetic instability. The chosen physical range is described in the main text.

\begin{figure*}
	\includegraphics[width=\linewidth]{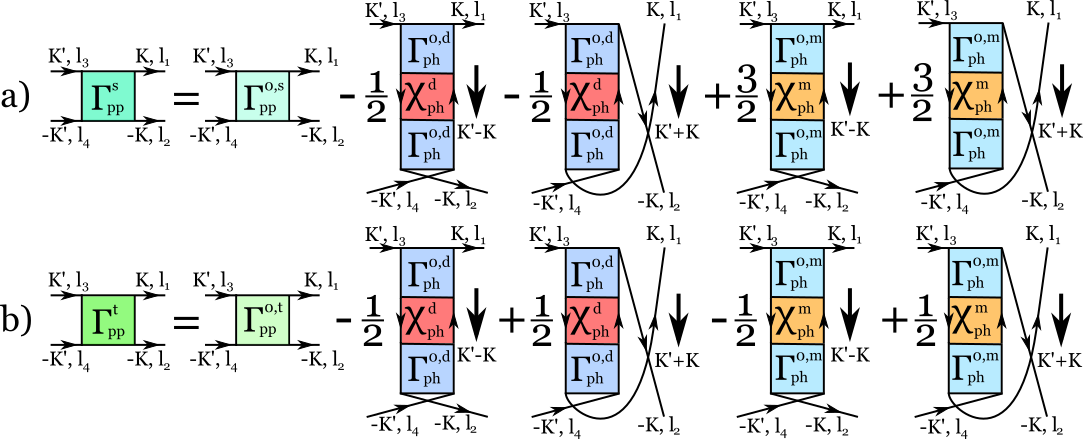}
	\caption{\label{fig:irr_vertex} Spin-diagonalized pairing vertices entering the particle-particle channel in a) singlet and b) triplet channels. The transfered four-momentum of the charge or spin fluctuation that are exchanged is shown on the side of each ladder function.}
\end{figure*}



\section{\label{sec:n-s_eliashberg}Normal-state Eliashberg Equation}

A superconducting transition to a singlet (triplet) state is signaled by an instability in the dressed susceptibility in p-p channel.
For pairs with vanishing center of mass frequency-momentum, the condition for a phase transition is given in the form of an eigenvalue problem known as the linearized Eliashberg equation
\begin{equation}
\begin{aligned}
	\label{eq:eliashberg}
	 -\left(\frac{k_BT}{N}\right)^2 \sum_{K'K''l_3...l_6} [\bm\Gamma^{s/t}_{pp}(0)]_{Kl_1l_2;K'l_3l_4} \times \\
	 [\bm\chi^0_{pp}(0)]_{K'l_3l_4;K''l_5l_6}\bm\Delta^{s/t}_{K''l_5l_6} = \lambda(T)\bm\Delta^{s/t}_{Kl_1l_2}.
\end{aligned}
\end{equation}

In spin- and charge-fluctuation theory, the pairing vertex for  singlet ($s$) and triplet ($t$) channels are given by \fref{fig:irr_vertex}.
The interaction is repulsive (attractive) for positive (negative) values of the pairing interaction. In the singlet channel, exchange of a charge fluctuation is attractive while a magnetic fluctuation is repulsive. In the triplet channel, exchange of any fluctuation with four-momentum $K'-K$ is attractive, while with $K'+K$ it is repulsive~\cite{S_nourafkan_correlation-enhanced_2016}.
 
From \eref{eq:eliashberg}, the eigenvectors $\bm\Delta$ with largest eigenvalue $\lambda$ are called leading gap functions.
Each irreducible representation of the system's group symmetry has a specific eigenvalue.
Using Arnoldi's algorithm, we obtain the leading gap functions in decreasing order of $\lambda$.

It is worth mentioning that the standard Eliashberg equation is usually solved in the superconducting phase, however \eref{eq:eliashberg} is written in the normal state at the vicinity of $T_C$. In other word, we are inspecting the pairing susceptibility divergence (also known as the Thouless criterion~\cite{S_thouless_perturbation_1960}) by approaching from the normal phase, hence we employ the name \textit{normal-state Eliashberg equation}. This method was widely used with great success in multi-band systems such as iron-based superconductors, for example see Ref~\citenum{S_graser_near-degeneracy_2009}.

\eref{eq:eliashberg} is a non-hermitian eigenvalue problem and the matrix involved has a very large size. Indeed, the pairing interaction depends on three Matsubara frequencies: two fermionic and one bosonic. The matrix indices are fermionic frequencies. The  \textit{center of mass} bosonic frequency is set to zero because we are looking for a thermodynamic instability. Since the pairing interaction decays very fast as a function of fermionic frequency (only electrons near the Fermi surface contribute to pairing), in most multi-band calculations the magnitude of the two fermionic frequencies are set equal and further dependence of the pairing interaction on fermionic frequency is neglected. In that case, summation over the fermionic Matsubara frequency in $\bm \chi^0_{pp}$ can be performed and the resulting equation is then similar to the one found in textbooks like Ref.~\citenum{S_abrikosov_methods_1975} for finding the pairing instability in the normal state. This is the so-called BCS approximation (see Eq. 28 of Ref.~\citenum{S_graser_near-degeneracy_2009} or Eq. 5 of Ref.~\citenum{S_scaffidi_pairing_2014}). Such an approach only captures even-frequency gap functions.

In our calculation, the full matrix structure in fermionic frequencies is kept. This allows us to capture both even- and odd- frequency gap functions. However, because of relatively narrow peaks in momentum space, we decided to keep only two possible values of fermionic frequencies and to consider the largest possible k-point grid. For some cases, we checked that increasing the number of fermionic Matsubara frequencies does not change the order of the leading gap function. Note that we have considered 1024 positive Matsubara frequencies in evaluating the bare particle-hole susceptibility, i.e, in Eq. (1) of the main text.

\section{\label{sec:prop}Properties of the gap function}

The gap function, entering the Eliashberg equation \eref{eq:eliashberg}, is a two-electrons condensate, or Cooper pair, written as $\bm\Delta(1,2)$, with the shortcut $1 \equiv ({\bf r}_1, \tau_1, l_1, \sigma_1)$ where position is ${\bf r}_1$,  imaginary time is $\tau_1$, orbital index $l_1$ and spin $\sigma_1$.
The two spatial (imaginary time) coordinates can be Fourier transformed, yielding a center of mass momentum ${\bf q}$ (bosonic frequency $\nu_n$) and  a relative momentum ${\bf k}$ (fermionic frequency $\omega_n$). With vanishing center of mass and thermodynamic equilibrium, we have $({\bf q}, i\nu_n)=0$ so that the elements of the gap function become $\bm\Delta_{Kl_1l_2}^{\sigma_1\sigma_2}$  with $K\equiv({\bf k},i\omega_n)$.

The normal state of SRO is invariant under any spacial transformation $g$ of the $D_{4h}$ point group.
An operator $\hat{g}$ of the group acts as follows
\begin{equation}
	\hat{g}\bm\Delta_K = G^T(g)\bm \Delta_{(R^{-1}(g){\bf k}, i\omega_n)} G(g)
\end{equation}
where $R(g) \in O(3)$ is a three-dimension rotation matrix and $G(g)$ is the direct product of two operators acting on spin and orbital spaces, respectively. 

Because we neglect spin-orbit coupling, these two contributions can be separated and spin is independent of momentum.
Spatially, the superconducting orders transform as different irreducible representations (irreps) of $D_{4h}$.
We use character theory to verify which irrep characterizes the symmetry of each gap function.

As explained in Ref.~\citenum{S_geilhufe_symmetry_2018}, the orthogonality of the irreps allows to construct the character projection operator $\hat{\mathcal{P}}^p$ of the irrep $p$ using  the character table of $D_{4h}$. Therefore, a gap function transforming as the irrep $q$ satisfies
\begin{equation}
	\label{eq:character}
	\hat{\mathcal{P}}^p \bm\Delta = \sum_{g\in D_{4h}} [\bar{\chi}^p(g)]^*\hat{g}\bm \Delta = \delta^{pq}\bm\Delta
\end{equation}
where $\bar{\chi}^p(g)$ is the character of transformation $g$ associated to irrep $p$.

Since spin is independent of momentum when spin-orbit coupling is neglected, the gap function can be spin-diagonalized into singlet (s) and triplet (t).
They are  odd and even solutions under spin exchange $\hat{S}$. Thereby
\begin{equation}
	\label{eq:spin_exchange}
	\hat{S}\bm\Delta^s = - \bm\Delta^{s}, \ \hat{S}\bm\Delta^t = \bm\Delta^t.
\end{equation}

Moreover, $D_{4h}$ being centrosymmetric, its irreps can be classified as even ($g$) or odd ($u$) under the parity operation $\hat{P}$, which reverses momentum ${\bf k}\rightarrow-{\bf k}$. The gap functions satisfy
\begin{equation}
	\label{eq:parity}
	\hat{P}\bm\Delta_{({\bf k}, i\omega)} = \pm \bm\Delta_{(-{\bf k}, i\omega)}.
\end{equation}

The gap function is not diagonal in orbital indices in general. It thus opens  the possibility to be even or odd under orbital exchange $\hat{O}$, that is 
\begin{equation}
	\label{eq:orb_exchange}
	[\hat{O}\bm\Delta_K]_{l_1l_2} = \pm \bm\Delta_{Kl_2l_1}.
\end{equation}

Let us define the operator $\hat{T}$ such that it transforms $i\omega_n$ into $-i\omega_n$. Like all other operators above, when we apply this operator twice, it is the identity, which means its eigenvalues are $\pm1$. Thus, we have

\begin{equation}
	\label{eq:rel_time_exchange}
	\hat{T} \bm \Delta_{({\bf k}, i\omega_n)} = \pm \bm \Delta_{({\bf k}, -i\omega_n)}.
\end{equation}

Given the definition of time-ordered product and the anticommutation of fermions, the identity $\bm\Delta(1,2) = -\bm\Delta(2,1)$ has to be satisfied. This implies the following constraint on the operators we just defined~\cite{S_linder_odd-frequency_2017}

\begin{equation}
	\hat{S}\hat{P}\hat{O}\hat{T} \bm \Delta(1,2) = -\bm\Delta(1,2).
\end{equation}

The gap function in general takes the form,
\begin{equation}
	\label{eq:orb_gap}
	\bm\Delta^{s/t}_K = \left( \begin{array}{ccc}	\bm\Delta_{xy;xy} 	& \bm\Delta_{xy;yz}	& \bm\Delta_{xy;xz} \\ 
										\bm\Delta_{yz;xy} 	& \bm\Delta_{yz;yz}	& \bm\Delta_{yz;xz} \\ 
										\bm\Delta_{xz;xy} 	& \bm\Delta_{xz;yz} 	& \bm\Delta_{xz;xz} 		\end{array} \right),
										\vspace{12pt}
\end{equation}
with $\bm\Delta_{l_1;l_2} \equiv \bm\Delta^{s/t}_{Kl_1l_2}$ an intra-orbital (inter-orbital) component if $l_1=l_2$ ($l_1\neq l_2$).
The relative contribution of all orbital components reveals in what orbital the Cooper pairs are mostly hosted.
A superconducting state that binds preferably an electron from orbital $l_1$ with one from orbital $l_2$ has dominant $\bm\Delta_{l_1l_2}$ and $\bm\Delta_{l_2l_1}$ components.
The gap function is said to be intra-orbital if the leading component has $l_1=l_2$ and inter-orbital otherwise.

\begin{figure}
	\includegraphics[width=\linewidth]{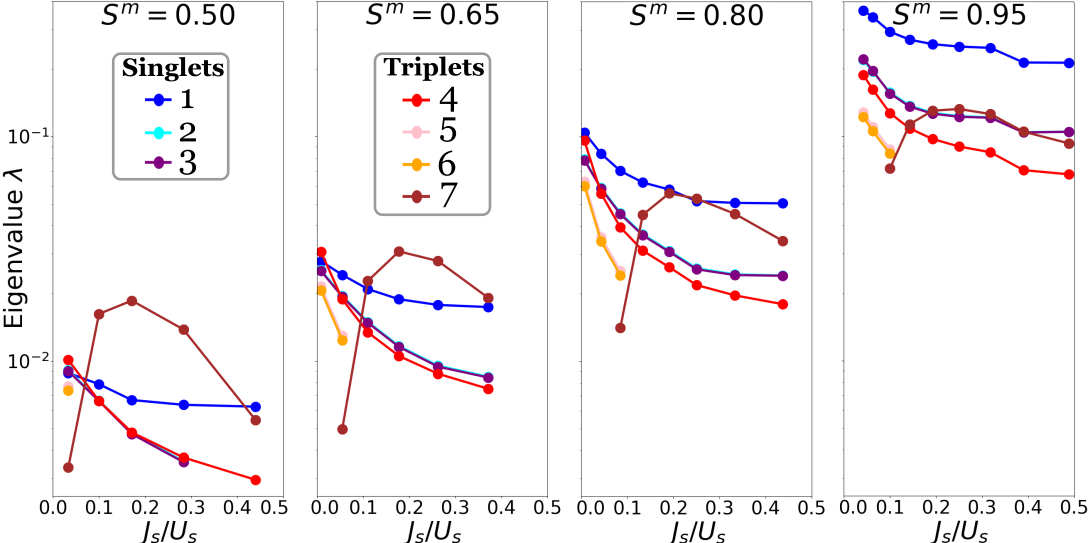}
	\caption{\label{fig:leading_eigs}Leading eigenvalues as a function of $J_s/U_s$ for various values of the Stoner factor $S^m$. The characteristics of each state are given in \tref{tab:states}.}
\end{figure}

\begin{table}[b]
	\begin{tabular*}{\linewidth}{ c c@{\extracolsep{\fill}}| c c c c c c c }
						& \quad	& \textbf{Dominant $\bm\Delta_{l_1;l_2}$}	& $\hat{S}$	& $\hat{P}$	& $\hat{O}$	& $\hat{T}$	& \textbf{S-Irrep}	& \textbf{Name} \\
		\hline
		\underline{\bf{1}}	&		&	$(xy, xy)$						&	-1		&	1		&	1		&	1		&	$B_{1g}$		& $d_{x^2-y^2}$ \\
		2				&		&	$(yz, yz), (xz, xz)$				&	-1		&	1		&	1		&	1		&	$B_{1g}$		& $d_{x^2-y^2}$ \\
		3				&		&	$(yz, yz), (xz, xz)$				&	-1		&	1		&	1		&	1		&	$A_{1g}$		& $s^{\pm}$ \\
		\underline{\bf{4}}	&		&	$(xy, xy)$						&	1		&	1		&	1		&	-1		&	$A_{1g}$		& $s$-wave \\
		5				&		&	$(yz, yz), (xz, xz)$				&	1		&	1		&	1		&	-1		&	$A_{1g}$		& $s$-wave \\
		6				&		&	$(yz, yz), (xz, xz)$				&	1		&	1		&	1		&	-1		&	$B_{1g}$		& $d_{x^2-y^2}$ \\
		\underline{\bf{7}}	&		&	$(xy, yz), (xy, xz)$				&	1		&	1		&	-1		&	1		&	$E_g$		& --- \\
	\end{tabular*}
	\caption{\label{tab:states}Properties of the leading gap functions of \fref{fig:leading_eigs}. Each column is described in the text.}
	\end{table}
	
\begin{figure}
		\includegraphics[width=\linewidth]{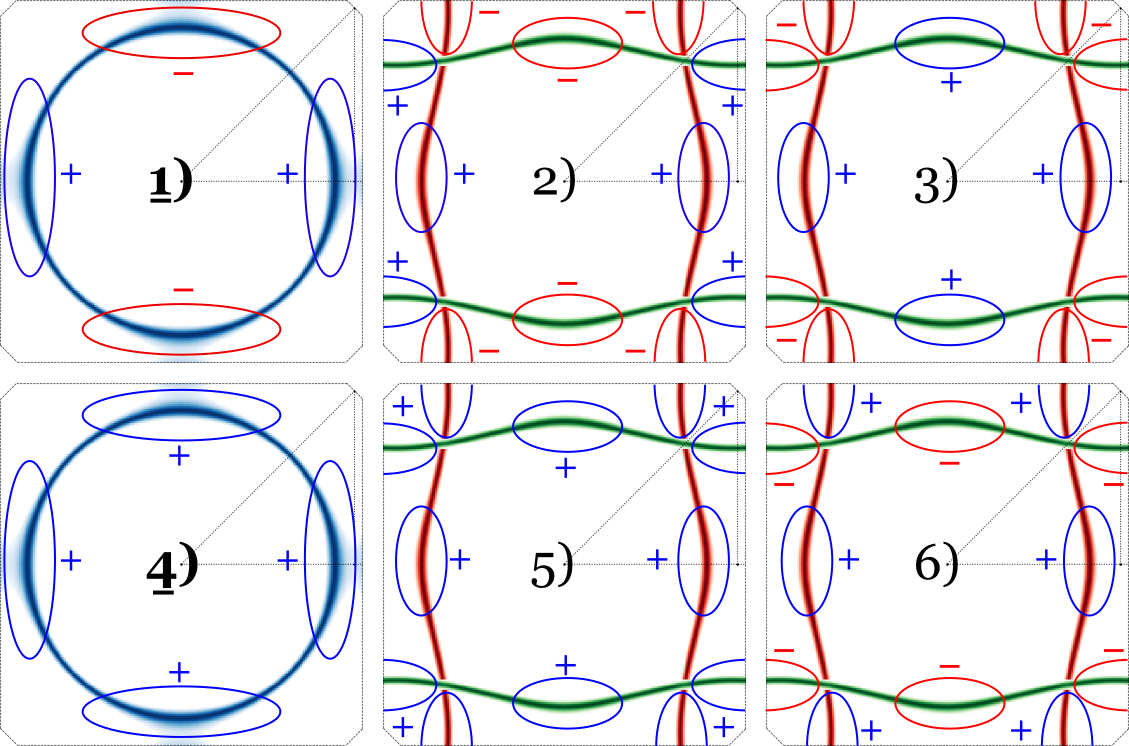}
		\caption{\label{fig:gap_schem}Schematics of the sign changing behavior of all dominant intra-orbital gap functions on the Fermi surface. The numbers correspond to those in \tref{tab:states}.}
\end{figure}

\section{Leading eigenvectors}\label{Sec:Eigenvectors}

The superconducting order parameters, or gap functions $\bm\Delta$, are the eigenvectors of the $\bm\Gamma_{pp}\bm\chi^0_{pp}$ matrices.
They are obtained, along with their corresponding eigenvalues $\lambda$, by solving the linearized Eliashberg equation \eref{eq:eliashberg}.
Each frequency-dependent gap function is an irreducible representation of the system's Shubnikov group of the second kind~\cite{S_geilhufe_symmetry_2018}.

For a given set of parameters $J_s/U_s$ and $S^m$, the gap function with largest eigenvalue is considered the leading instability.
\fref{fig:leading_eigs} presents a few leading eigenvalues in both spin-singlet and spin-triplet channels. We only keep the symmetries that are important in each range of parameters.
Each label corresponds to a distinct symmetry, with specific properties explicitly given in \tref{tab:states}. States number 2 and 3 (5 and 6) are almost degenerate, so they are difficult to distinguish. State number 7 is doubly degenerate.

The first column of \tref{tab:states} is the dominant orbital components $l_1;l_2$ of the gap function, as written in \eref{eq:orb_gap}.
The second to fifth columns show whether the gap function is even or odd under spin exchange $\hat{S}$, parity $\hat{P}$, orbital exchange $\hat{O}$ and relative time exchange $\hat{T}$.
The effect of these operators are given by Eqs. (\ref{eq:spin_exchange}) to (\ref{eq:rel_time_exchange}).
The sixth column gives the irreducible representation of the spatial group (S-Irrep), as obtained using \eref{eq:character}. 

The last column gives the symmetry name including angular momentum. To help understand the spatial structure of these gaps, \fref{fig:gap_schem} shows the sign-changing behavior of intra-orbital gap functions on the FS. Gap 7 is not showed because it pairs electrons on different orbitals, which is difficult to represent.

For each set of parameters studied, the eigenvector with largest eigenvalue was used to construct the phase diagram of Fig.~4 in the main text.
The corresponding states have bold and underlined labels in \tref{tab:states}.

\section{Odd-orbital pairing mechanism}
To discuss inter-orbital pairing, we must focus on the inter-orbital pairing interactions, i.e., $\left[ \bm \Gamma^t_{pp}(0) \right]_{Klm;K'lm}$ and $\left[ \bm \Gamma^t_{pp}(0) \right]_{Klm;K'ml}$ with $l \neq m$. Keeping only the largest component of the pairing susceptibility, the former connect inter-orbital gap function $\bm \Delta_{lm}$ to itself while the latter connect $ \bm \Delta_{lm}$ to $\bm \Delta_{ml}$. We will show that around $J_s/U_s \sim 0.25$, $\left[ \bm \Gamma^t_{pp}(0) \right]_{Klm;K'lm}$ is attractive while $\left[ \bm \Gamma^t_{pp}(0) \right]_{Klm;K'ml}$ is repulsive, hence promoting an odd-orbital gap function. 

As shown in \fref{fig:irr_vertex}, the pairing in the triplet channel is
\begin{widetext}
\begin{align}
	\left[ \bm \Gamma^t_{pp}(0) \right]_{Kl_1l_2;K'l_3l_4} = \left[ \bm \Gamma^{0,t}_{pp} \right]_{l_1l_2;l_3l_4} & - \frac{1}{2} \left[\bm \Phi^d_{ph}(K'-K)\right]_{l_2l_4;l_3l_1} +\frac{1}{2}\left[ \bm \Phi^d_{ph}(K'+K)\right]_{l_1l_4;l_3l_2}  \nonumber \\
		& - \frac{1}{2} \left[ \bm \Phi^m_{ph}(K'-K)\right]_{l_2l_4;l_3l_1} + \frac{1}{2} \left[ \bm \Phi^m_{ph}(K'+K)\right]_{l_1l_4;l_3l_2}
	\label{eq:triplet_pairing}
\end{align}
where $\bm \Phi^{d(m)}_{ph}$ are the density (magnetic) ladder functions. They are given by
\begin{equation}
	\left[ \bm \Phi^{d(m)}_{ph}(Q)\right]_{l_1l_2;l_3l_4} = \sum_{l_5...l_8} \left[ \bm \Gamma^{0,d(m)}_{ph}\right]_{l_1l_2;l_5l_6} \left[ \bm \chi^{d(m)}_{ph}(Q)\right]_{l_5l_6;l_7l_8} \left[ \bm \Gamma^{0,d(m)}_{ph} \right]_{l_7l_8;l_3l_4}.~\label{eq:ladder}
\end{equation}
Since the dominant components of the ladder function are positive, the pairing interaction in the triplet channel, \eref{eq:triplet_pairing}, has both attractive and repulsive components.

In the RPA approximation, the particle-hole irreducible vertex function appearing in the ladder functions is defined as \eref{eq:irr_ph_vertex}.
Focusing on the inter-orbital component of the pairing interaction $\left[ \bm \Gamma^t_{pp}(0) \right]_{Klm;K'lm}$ with $l\neq m$, one can see from \eref{eq:triplet_pairing} that it can be written in terms of the $mm;ll$ and $lm;lm$ components of the ladder functions. The $mm;ll$ component has a $(-1/2)$ multiplicative factor in \eref{eq:triplet_pairing}, hence leads to an attractive pairing interaction while the $lm;lm$ component gives a repulsive one. The role of these components reverse if we consider $\left[ \bm \Gamma^t_{pp}(0) \right]_{Klm;K'ml}$ instead.

Using~\eref{eq:ladder} and the restrictions imposed by the irreducible vertex function, \eref{eq:irr_ph_vertex}, the dominant terms contributing in the $mm;ll$ and $lm;lm$ components of the ladder functions are
\begin{align}
	\left[ \bm \Phi^{d/m}_{ph}(K'-K)\right]_{mm;ll} &\simeq	  \left[ \bm \Gamma^{0,d/m}_{ph}\right]_{mm;mm} \left[ \bm \chi^{d/m}_{ph}(K'-K)\right]_{mm;mm} \left[ \bm \Gamma^{0,d/m}_{ph} \right]_{mm;ll}\nonumber\\
		&+ \left[ \bm \Gamma^{0,d/m}_{ph}\right]_{mm;ll} \left[ \bm \chi^{d/m}_{ph}(K'-K)\right]_{ll;ll} \left[ \bm \Gamma^{0,d/m}_{ph} \right]_{ll;ll}, \\
	\left[ \bm \Phi^{d/m}_{ph}(K'+K)\right]_{lm;lm} &\simeq \left[ \bm \Gamma^{0,d/m}_{ph}\right]_{lm;lm} \left[ \bm \chi^{d/m}_{ph}(K'+K)\right]_{lm;lm} \left[ \bm \Gamma^{0,d/m}_{ph} \right]_{lm;lm}\nonumber\\
		&+ \left[ \bm \Gamma^{0,d/m}_{ph}\right]_{lm;ml} \left[ \bm \chi^{d/m}_{ph}(K'+K)\right]_{ml;ml} \left[ \bm \Gamma^{0,d/m}_{ph} \right]_{ml;lm},
\end{align}
where we only kept the dominant components of the susceptibilities, $[\bm \chi^{d/m}_{ph}]_{l_1l_2;l_1l_2}$. Using the explicit expressions on the right-hand side of \eref{eq:irr_ph_vertex} we find, 
\begin{align}
	\left[ \bm \Phi^{d}_{ph}(K'-K)\right]_{mm;ll} &\simeq U_s(2U'_s - J_s) \left(\left[ \bm \chi^{d}_{ph}(K'-K)\right]_{mm;mm} +
	\left[ \bm \chi^{d}_{ph}(K'-K)\right]_{ll;ll} \right),\\
	\left[ \bm \Phi^{m}_{ph}(K'-K)\right]_{mm;ll} &\simeq	  U_sJ_s \left(\left[ \bm \chi^{m}_{ph}(K'-K)\right]_{mm;mm} +
	\left[ \bm \chi^{m}_{ph}(K'-K)\right]_{ll;ll} \right),\\
	\left[ \bm \Phi^{d}_{ph}(K'+K)\right]_{lm;lm} &\simeq (-U'_s + 2J_s)^2 \left[ \bm \chi^{d}_{ph}(K'+K)\right]_{lm;lm} +(J_s)^2 \left[ \bm \chi^{d}_{ph}(K'+K)\right]_{ml;ml} ,\\
	\left[ \bm \Phi^{m}_{ph}(K'+K)\right]_{lm;lm} &\simeq (U'_s)^2 \left[ \bm \chi^{m}_{ph}(K'+K)\right]_{lm;lm} +(J_s)^2 \left[ \bm \chi^{m}_{ph}(K'+K)\right]_{ml;ml}.
\end{align}
\end{widetext}

One can see that increasing the Hund's coupling suppresses the charge fluctuations as expected: At $J_s=0$, intra- and inter-orbital density-density interactions are equal $U_s=U'_s$, hence electrons can hop locally between different orbitals without any extra energy cost. This degeneracy is lifted for a finite $J_s$, decreasing the charge (orbital) fluctuations. On the other hand, a larger Hund's coupling increases the spin-fluctuations as seen from the last equation. Since, the inter-orbital pairing interaction is given by summation of the charge and spin fluctuation (see Eq.~1), the role of the Hund's coupling is not trivial. Nevertheless, assuming $U'_s=U_s-2J_s$, then for $J_s/U_s \sim 0.25$, one can see that the $mm;ll$ component of the ladder function is larger than the $lm;lm$ component. Indeed, for $J_s/U_s=0.25$, the coefficient $(-U'_s + 2J_s)$ vanishes while $U_s(2U'_s - J_s)$ is large.

A similar analysis can be done for $\left[ \bm \Gamma^t_{pp}(0) \right]_{Klm;K'ml}$. The only difference is that $mm;ll$ component of the ladder function appears with positive coefficient while $lm;lm$ has negative coefficient leading to a repulsive pairing interaction. Since, $\left[ \bm \Gamma^t_{pp}(0) \right]_{Klm;K'ml}$ connect $\Delta_{lm}$ to $\Delta_{ml}$, it leads to a odd-inter-orbital pairing.

\end{document}